\documentclass[10pt,aps,prl,amsfonts,amsmath,amssymb,raggedbottom,longbibliography,reprint,superscriptaddress,citeautoscript]{revtex4-2}

\usepackage{amsmath}
\usepackage{graphicx}
\graphicspath{{Figures/}{Figs/}}

\date{\today}

\begin{document}











\title{Ultrafast lattice disordering can be accelerated by electronic collisional forces}

\author{Gilberto A. de la Pe\~na Mu\~noz}
\affiliation{Stanford PULSE Institute, SLAC National Accelerator Laboratory, Menlo Park, CA 94025, USA.}
\affiliation{Stanford Institute for Materials and Energy Sciences, SLAC National Accelerator Laboratory, Menlo Park, CA 94025, USA.}

\author{Alfredo A. Correa}
\email{correaa@llnl.gov}
\affiliation{Quantum Simulations Group, Lawrence Livermore National Laboratory, Livermore, California 94551, USA.}

\author{Shan Yang}
\affiliation{Department of Mechanical Engineering and Materials Science, Duke University, Durham, NC 27708, USA.}

\author{Olivier Delaire}
\affiliation{Department of Mechanical Engineering and Materials Science, Duke University, Durham, NC 27708, USA.}

\author{Yijing Huang}
\affiliation{Stanford PULSE Institute, SLAC National Accelerator Laboratory, Menlo Park, CA 94025, USA.}
\affiliation{Stanford Institute for Materials and Energy Sciences, SLAC National Accelerator Laboratory, Menlo Park, CA 94025, USA.}

\author{Allan S. Johnson}
\affiliation{ICFO-Institut de Ciencies Fotoniques, The Barcelona Institute of Science and Technology, Av. Carl Friedrich Gauss 3, 08860 Castelldefels (Barcelona), Spain.}
\affiliation{IMDEA Nanoscience, Calle Faraday 9, 28049, Madrid, Spain}

\author{Tetsuo Katayama}
\affiliation{Japan Synchrotron Radiation Research Institute, 1-1-1 Kouto, Sayo-cho, Sayo-gun, Hyogo 679-5198, Japan.}
\affiliation{RIKEN SPring-8 Center, 1-1-1 Kouto, Sayo, Hyogo 679-5148, Japan.}

\author{Viktor Krapivin}
\affiliation{Stanford PULSE Institute, SLAC National Accelerator Laboratory, Menlo Park, CA 94025, USA.}
\affiliation{Stanford Institute for Materials and Energy Sciences, SLAC National Accelerator Laboratory, Menlo Park, CA 94025, USA.}

\author{Ernest Pastor}
\affiliation{ICFO-Institut de Ciencies Fotoniques, The Barcelona Institute of Science and Technology, Av. Carl Friedrich Gauss 3, 08860 Castelldefels (Barcelona), Spain.}
\affiliation{IPR–Institut de Physique de Rennes, CNRS-Centre national de la recherche scientifique, UMR 6251 Universit\'{e} de Rennes, 35000 Rennes, France}

\author{David A. Reis}
\affiliation{Stanford PULSE Institute, SLAC National Accelerator Laboratory, Menlo Park, CA 94025, USA.}
\affiliation{Stanford Institute for Materials and Energy Sciences, SLAC National Accelerator Laboratory, Menlo Park, CA 94025, USA.}

\author{Samuel Teitelbaum}
\affiliation{Physics Department, Arizona State University, Phoenix, Arizona, USA.}
\author{Luciana Vidas}
\affiliation{ICFO-Institut de Ciencies Fotoniques, The Barcelona Institute of Science and Technology, Av. Carl Friedrich Gauss 3, 08860 Castelldefels (Barcelona), Spain.}

\author{Simon Wall}
\affiliation{Department of Physics and Astronomy, Aarhus University, Aarhus, Denmark.}

\author{Mariano Trigo}
\email{mtrigo@slac.stanford.edu}
\affiliation{Stanford PULSE Institute, SLAC National Accelerator Laboratory, Menlo Park, CA 94025, USA.}
\affiliation{Stanford Institute for Materials and Energy Sciences, SLAC National Accelerator Laboratory, Menlo Park, CA 94025, USA.}


\begin{abstract}
\textbf{
In the prevalent picture of ultrafast structural phase transitions, the atomic motion occurs in a slowly varying potential energy surface determined adiabatically by the fast electrons. However, this ignores non-conservative forces caused by electron-lattice collisions, which can significantly influence atomic motion. Most ultrafast techniques only probe the average structure and are less sensitive to random displacements, and therefore do not detect the role played by non-conservative forces in phase transitions. Here we show that the lattice dynamics of the prototypical insulator-to-metal transition of \(\mathrm{VO_2}\) cannot be described by a potential energy alone. We use the sample temperature to control the preexisting lattice disorder before ultrafast photoexcitation across the phase transition and our ultrafast diffuse scattering experiments show that the fluctuations characteristic of the rutile metal develop equally fast (\(120~\mathrm{fs}\)) at initial temperatures of 100 K and 300 K. This indicates that additional non-conservative forces are responsible for the increased lattice disorder. These results highlight the need for more sophisticated descriptions of ultrafast phenomena beyond the Born-Oppenheimer approximation as well as ultrafast probes of spatial fluctuations beyond the average unit cell measured by diffraction.
}
\end{abstract}

\maketitle

\section{Main text}
Ultrafast structural transformations are generally described in terms of atoms moving in potential energy surfaces that are determined adiabatically by the motion of the electrons on much faster timescales.
This well-known Born-Oppenheimer approximation is the usual description of nuclear motion in chemical dynamics, coherent lattice vibrations~\cite{Fritz2007,Zeiger1992}, and ultrafast structural transformations in solids~\cite{Baum2007,Sciaini2009,Lindenberg2005,Hillyard2007,Morrison2014,Huber2014,Beaud2014,Wall2018,Li2022}.
This description ignores the bi-directional, non-adiabatic~\cite{Head-Gordon1995} exchange of energy between the ions and the electrons~\cite{Duffy2006,Darkins2018,Darkins2018simulating}. 
The approximation breaks down when there is no clear separation between the time- (or energy-) scale of the atomic and electronic motions such as in metals or across IMTs.
In solids, this breakdown could manifest as enhanced lattice fluctuations caused by random electron-lattice collisions and an effective electronic friction~\cite{Head-Gordon1995}. 
However, it is difficult to observe these lattice fluctuations by ultrafast diffraction because it is primarily sensitive to \emph{average} structural changes.
The current framework of time-dependent potentials~\cite{Sun2020,Beaud2014,Wall2018} seems to describe the motion of the average structure captured by ultrafast diffraction experiments, but new experiments are required to probe the fluctuations caused by the non-conservative forces.


\begin{figure*}[htb]
\centering
\includegraphics[width=0.8\textwidth]{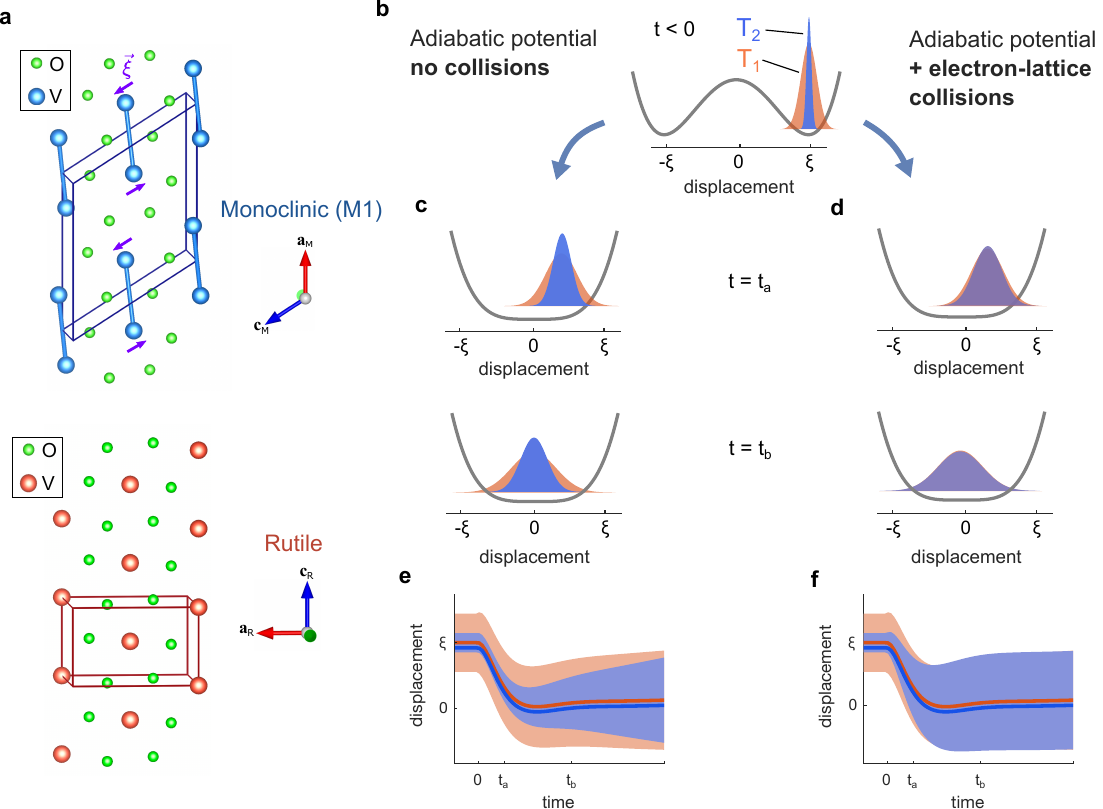} 
\caption{
    \textbf{Effect of non-conservative electronic forces on structural dynamics.}
    (a) Schematic structure of \(\mathrm{VO_2}\).
    Light-blue (orange) spheres represent vanadium atoms in the monoclinic M1 (rutile R) phase and green spheres are the oxygen atoms.
    Some equivalent sets of vanadium atoms are omitted for visual clarity.
    The M1 and rutile unit cells are shown with blue and red lines, respectively. 
    (b) Schematic of the potential surface  at \(t < 0\) (gray lines) and distribution of vanadium displacements around the M1 order parameter \(\vec{\xi}\) for two initial temperatures \(T_1 > T_2\). 
    At \(t > 0\) (c-d) photoexcitation changes the potential energy and the distribution shifts towards \(\vec{\xi} = 0\) and broadens, a signature of increased disorder.
    (c) and (d) correspond to scenario without, and with, electron-lattice collisions, respectively.
    (e) and (f) show a schematic of the time evolution of the mean (solid curves) and width (shaded area) of the displacement distributions without (e) and with (f) collisions for the two temperatures. 
    The solid lines are slightly displaced vertically for clarity.
    Electron-lattice collisions increase the disorder (variance) in (d) and (f) faster than in (c) and (e).
    \label{fig:1}
}
\end{figure*}

Here we overcome this limitation by simultaneously measuring the ultrafast x-ray diffraction and diffuse scattering using an x-ray free electron laser.
Using these measurements we demonstrate the effect of the non-adiabatic forces on an insulator-metal transition on a single-crystal of \(\mathrm{VO_2}\).
Femtosecond x-ray diffraction following above-gap photoexcitation tracks the evolution of the Bragg peaks of the equilibrium M1 phase, which reflect the long-range-order (LRO) of the vanadium dimers that double the unit cell~\cite{Comes1974}. Importantly, their response directly reflects the shape of the photoexcited potential\cite{Wall2018}.
More importantly, we simultaneously record the diffuse intensity between Bragg peaks, a signature of spatial lattice fluctuations and disorder, particularly in the vanadium dimers~\cite{Budai2014}.
We change the equilibrium sample temperature between \(300\) and \(100~\mathrm{K}\) to control the atomic velocities prior to photo-excitation.
We observe that both the peaks and especially the diffuse intensity respond with virtually the same timescale between the two sample temperatures. 
This is surprising because without additional forces, the soft, flat potential surface\cite{Wall2018} cannot accelerate the disorder and the diffuse intensity cannot increase faster than dictated by the initial velocities. Without additional sources of disorder the intensity onset should slow down at lower temperature, as predicted by ab-initio simulations of adiabatic potentials (Extended Data Fig. 1) and as observed in other photoinduced order-disorder transitions~\cite{Wang2020}.
We conclude that non-adiabatic forces from electronic collisions with hot electrons~\cite{Stern2018,Chase2016} are as important as the potential~\cite{Wall2018} in the dynamics of the IMT in \(\mathrm{VO_2}\). The observed timescale of $120~\mathrm{fs}$, fast compared with other metals\cite{Chase2016}, is indicative of an unusually strong electron-lattice coupling in \(\mathrm{VO_2}\). 
While these qualitative conclusions are drawn from general arguments, we illustrate the behavior of the fluctuations semi-quantitatively with a simple model that includes disorder.



On cooling, \(\mathrm{VO_2}\) transforms at \(T_\text{c} = 340~\mathrm{K}\) from a metal~\cite{Berglund1969,Goodenough1971,Zylbersztejn1975} with a tetragonal rutile (R) structure~\cite{Mcwhan1974} to a monoclinic (M1) insulating phase (Fig.~\ref{fig:1}a). 
At \(T < T_\text{c}\) the system exhibits a frozen distortion represented by \(\vec{\xi}\) in Fig.~\ref{fig:1}a and new Bragg peaks at wavevectors \((0, 1/2, 1/2)_\text{R}\) in the rutile Brillouin zone~\cite{Wall2018}.
Ultrafast experiments have provided valuable information on the transformation pathway between the M1 and R phases.
The salient observations are the fast \(100~\mathrm{fs}\) loss of LRO in the vanadium dimers observed by ultrafast diffraction, concomitant with a sharp change in optical conductivity hallmark of a transient metal~\cite{Cavalleri2004,Cavalleri2001,Baum2007,Morrison2014,Otto2019,Li2022}.
However, diffraction primarily probes the average order parameter \(\vec{\xi}\), which lacks sensitivity to spatial fluctuations caused by the non-conservative forces.

Recent diffuse and inelastic x-ray scattering measurements in equilibrium showed a direct connection between the soft transverse acoustic (TA) phonons and the enhanced lattice fluctuations responsible for the high entropy of the metallic phase~\cite{Budai2014}.
Previous work using ultrafast diffuse scattering~\cite{Wall2018} found unexpectedly fast lattice disordering upon photoexcitation, a consequence of the order-disorder nature of this transition.
Furthermore, ab-initio calculations~\cite{Budai2014,Wall2018,Li2022} show that the rutile and the photoexcited states are characterized by a nearly flat, anharmonic potential along the coordinate \(\vec{\xi}\), which enables the large variance of displacements and the high lattice entropy.
Furthermore, this potential describes the diffuse intensity\cite{Budai2014} and the dynamics of the average structure extremely well\cite{Wall2018}.
These results underscore the importance of lattice fluctuations in both the thermodynamic equilibrium and the transformation pathway as well as the need for probes of fluctuations such as diffuse scattering.

Fig.~\ref{fig:1}b-f shows schematically the generic behavior of the atomic displacement distribution across a photoinduced structural phase transition. We consider two idealized but well distinguished scenarios:
In one case the dynamics are primarily governed by a slowly varying adiabatic potential (c, e), and in the other scenario there are additional non-conservative forces due to electron-lattice collisions (d, f). 
The distinction is that the width of the distributions, visible as diffuse intensity, increases faster in (d) and (f) due to additional non-adiabatic forces.
In the context of VO$_2$ the distributions in Fig.~\ref{fig:1} represent the vanadium positions relative to the rutile structure (distributed over multiple unit cells) and their changes due to photoexcitation.
Initially at \(t < 0\) the system is dimerized (gray potential curve in Fig.~\ref{fig:1}b) with a displacements distribution centered at \(\vec{\xi}\) and initial width dictated by the initial temperature, \(T_1\) and \(T_2\), respectively. 
The wide photoinduced potential at \(t>0\) enables the distributions to broaden resulting in enhanced spatial fluctuations (Figs.~\ref{fig:1}c and \ref{fig:1}d).
The solid line and shaded areas in Figs.~\ref{fig:1}e and \ref{fig:1}f depict the dynamics of the mean and variance for the two scenarios, respectively.
In the scenario without additional collisions (e), the potential cannot accelerate the disorder and the increase in the variance and diffuse intensity are primarily determined by the initial temperature and become slower at lower temperature.
Such is the case of the ultrafast melting in InSb~\cite{Lindenberg2005,Gaffney2005,Hillyard2007} where the loss of LRO is slower for a lower initial temperature~\cite{Wang2020}.
However, when the electron-lattice collisions are strong (Fig. \ref{fig:1}f) they can dominate the increase in the disorder, which becomes nearly unaffected by the initial temperature. 
Importantly, diffraction is only sensitive to the mean (solid lines in Figs \ref{fig:1}e-f) and insensitive to disorder, thus these two scenarios are indistinguishable from a diffraction experiment alone.
Here we use x-ray diffuse scattering to show that \(\mathrm{VO_2}\) behaves as in the case in Fig.~\ref{fig:1}f and Fig.~\ref{fig:1}d, indicating that the photoexcited electrons cannot be ignored in describing the dynamics.

\begin{figure}[htb]
\centering 
\includegraphics[width=\columnwidth]{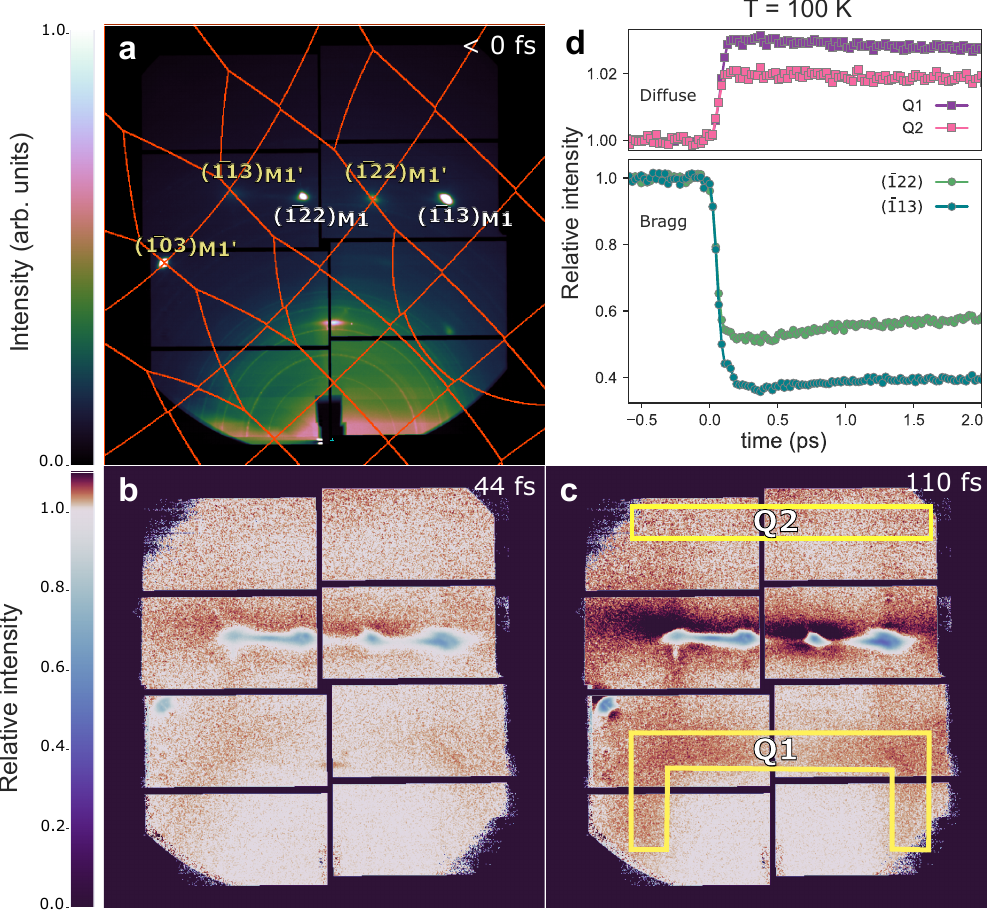}
\caption[]{\textbf{Ultrafast x-ray diffraction and diffuse scattering from \(\mathrm{VO_2}\).}
    (a) Representative static x-ray intensity captured with \(12~\mathrm{keV}\) photons at a base temperature of \(100~\mathrm{K}\).
    The bright spots originate from wavevectors near Bragg points, which are labeled by the corresponding Miller indices for domains M1 and M1' (see text for details).
    Solid red lines represent the Brillouin zone (BZ) edges for the M1 structure;
    BZs for M1' are omitted for clarity.
    (b-c) Relative diffuse intensity at representative delays of \(44\) and \(110~\mathrm{fs}\) after photoexcitation with a near-infrared laser pulse centered at \(800~\mathrm{nm}\) with \(50 ~\mathrm{fs}\) pulse duration.
    The regions of reciprocal space labeled \(Q_1\) and \(Q_2\) are characteristic of the broken dimers~\cite{Wall2018}.
    (d) Integrated intensity as a function of time for representative Bragg peaks and for diffuse regions \(Q_1\) and \(Q_2\).
    \label{fig:2}
}
\end{figure}

Fig.~\ref{fig:2}a shows a representative static x-ray scattering image of the M1 phase at \(100~\mathrm{K}\) taken with \(12~\mathrm{keV}\) x-ray photons at the SACLA x-ray free electron (XFEL) facility in Japan~\cite{Ishikawa2011} (see Methods for further experimental details).
The bright spots correspond to points near the zone-center of the M1 Brillouin zone (BZ) where the intensity is primarily from the structure factor of the nearest M1 Bragg point.
In this geometry, we observe contributions from two domains of the M1 structure which we label M1 and M1' (see Extended Data Fig. 2 for BZs for M1'), where M1' is rotated from M1 by \(90^\circ\) around the rutile c-axis.
The red lines in Fig.~\ref{fig:2}a represent the boundaries of the BZs for the M1 domain.
In Fig.~\ref{fig:2}b-c we show representative snapshots of the relative intensity, $I(t)/I(t < 0)$, where $I(t)$ is the intensity at time $t$ and $I(t < 0)$ is the equilibrium intensity before the pump pulse.
There are appreciable changes in the diffuse intensity at \(t = 44~\mathrm{fs}\) which develop almost fully at \(110~\mathrm{fs}\), with little change up to several tens of \(\mathrm{ps}\)~\cite{Wall2018}.
The sample returns to equilibrium after several nanoseconds, shorter than the \(40~\mathrm{ms}\) repetition period of the pump pulses.
Fig.~\ref{fig:2}d shows the evolution of the \((\bar1\,2\,2)_\text{M1}\) and \((\bar1\,1\,3)_\text{M1}\) structure factors at \(T = 100~\mathrm{K}\) (lower panel) together with the diffuse intensity integrated over the regions labeled \(Q_1\) and \(Q_2\) in Fig.~\ref{fig:2}c (top panel). 
The diffuse intensity is characteristic of the  disordered vanadium dimers and strongly resembles the equilibrium pattern of the rutile phase~\cite{Wall2018}.

\begin{figure}[htb]
\centering 
\includegraphics[width=0.4\textwidth]{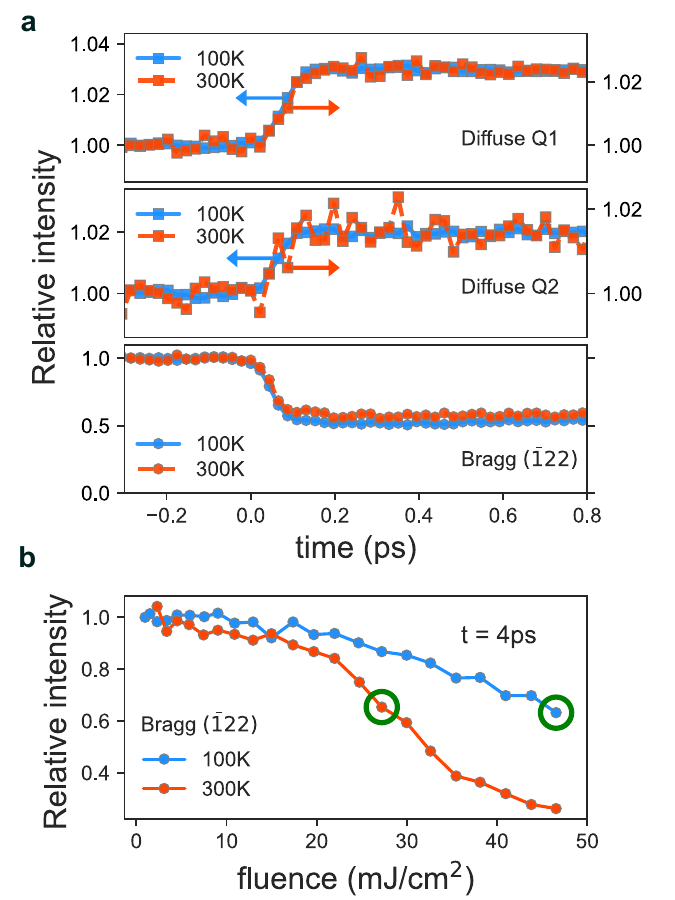} 
\caption{\textbf{Influence of the initial temperature on the dynamics.}
    (a) Comparison of the dynamics of the relative intensity of the \((\bar{1}\,2\,2)\) peak and the diffuse intensity in regions \(Q_1\) and \(Q_2\) between \(100\) and \(300~\mathrm{K}\) at incident fluences of \(46.5~\mathrm{mJ/cm^2}\) and \(28~\mathrm{mJ/cm^2}\), respectively, indicated by the green circles in (b).
    (b) Incident fluence dependence of the relative intensity of \((\bar{1}22)\) at \(t = 4~\mathrm{ps}\) for \(100\) and \(300~\mathrm{K}\). 
    \label{fig:3}
}
\end{figure}

Fig.~\ref{fig:3} (a) shows the evolution of the relative intensity of the \((\bar1\,2\,2)_\text{M1}\) Bragg peak (lower panel) at initial temperatures of \(100~\mathrm{K}\) and \(300~\mathrm{K}\). 
The timescale for the intensity drop of \(100~\mathrm{fs}\) is indistinguishable between the two temperatures (see Extended Data Fig.~3 for timescale definition and fits). 
We also show the integrated dynamics of the relative intensity in the \(Q_1\) region at the two temperatures (top panel). 
Here too, the intensity increases with virtually identical timescale of \(120~\mathrm{fs}\) between the two temperatures.
There is also no appreciable difference in the timescale of the \(Q_2\) region between the two temperatures (middle panel).
These timescales are well above the experimental resolution \(50~\mathrm{fs}\), and are also consistent with prior measurements at a different XFEL facility~\cite{Wall2018}.
Fig. \ref{fig:3} (b) shows the fluence dependence of the relative intensity at \(4~\mathrm{ps}\) for \(300~\mathrm{K}\) and \(100~\mathrm{K}\). 
This shows that the material can be transformed at both initial temperatures, and is consistent with ultrafast reflectivity measurements\cite{Vidas2020}.
The results of the \(Q_1\) and \(Q_2\) regions in Fig.~\ref{fig:3} (a) imply that the preexisting fluctuations at $t < 0$, controlled by the initial temperature, do not affect the disordering initiated by photoexcitation and that the system quickly loses memory of the initial thermal distribution.
This contrasts sharply with other order-disorder transitions such as ultrafast melting of InSb where the initial melting dynamics are inertial and temperature dependent~\cite{Wang2020}.
It also contrasts strongly with the slower disorder at lower temperature predicted by ab-initio molecular dynamics (AIMD) (see Extended Data Fig. 1), which only considers forces in the Born-Oppenheimer approximation.
The fast disordering at low temperatures observed here implies that additional fluctuating forces not captured by the potential, which otherwise describes the quasi-equilibrium diffuse intensity, phonon dispersion\cite{Budai2014} and Bragg peaks dynamics quite well~\cite{Wall2018,Li2022}, must be affecting the lattice.

\begin{figure*}[htb]
\includegraphics[width=0.95\textwidth]{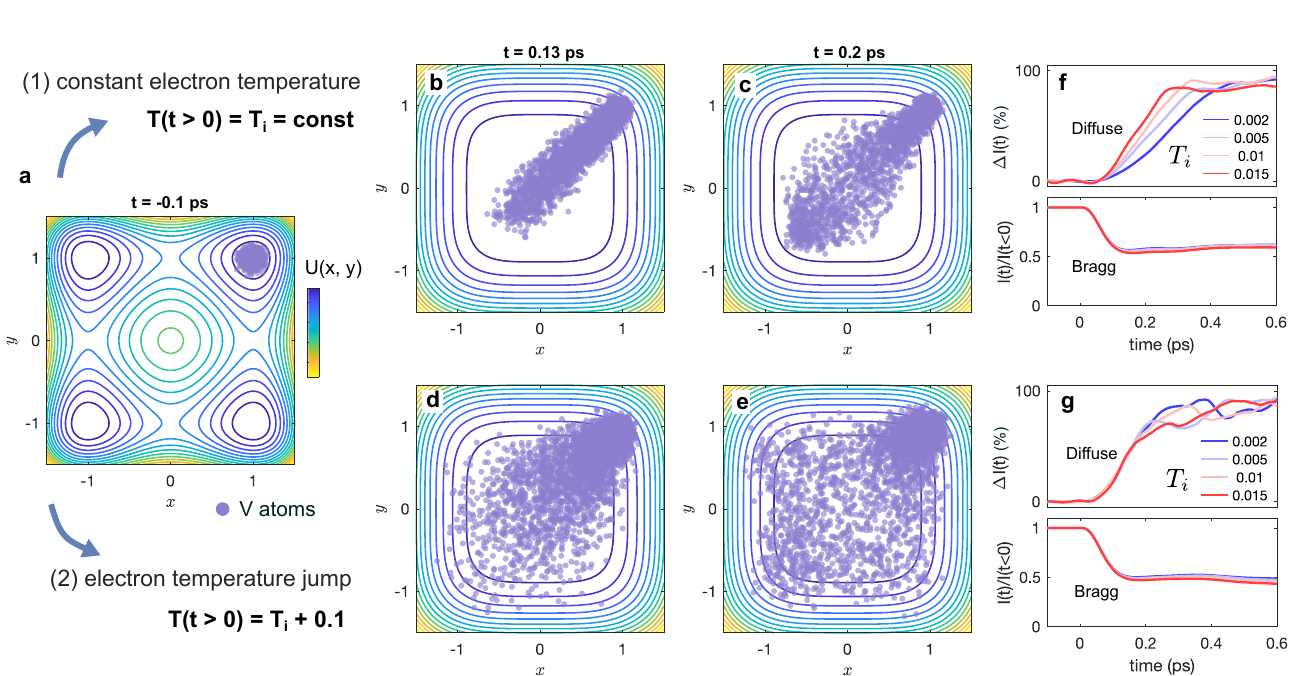} 
\caption[]{
    {\bf Model of the disorder with non-conservative forces.}
    (a) Contour plot of the potential \(U(x, y, t< 0)\) and \(U(x, y, t > 0)\) (b - e), respectively.
    The dots in (a) represent the vanadium atoms in the initial thermal distribution at \(T = 0.005\) and out-of-equilibrium in (b-e).
    Top and bottom rows represent two idealized scenarios from Eq. (\ref{eq:eom}): 1) constant bath temperature (b-c) and (f), and 2) bath temperature jump (d-e) and (g). All trajectories in (b-e) start with the thermal distribution in (a) at \(T = 0.005\).
    (f-g) Relative intensity of 2D Bragg peaks \((2.5, 2.5)\) (in reciprocal lattice units of the 2D lattice) representative of the dimerized doubled unit cell (bottom panels) and the change in diffuse intensity, $\Delta I(t)$ integrated along wavevectors \((0.1, 0.1) - (0.45, 0.45)\) (top panels). 
    For easier comparison between temperatures, $\Delta I(t)$ curves in (f, g) are shown as percentage of their respective saturation value at $t > 2$~ps.
    The legends in (f-g) show the temperature of the initial distribution \(T(t < 0)\) in units of the potential well depth. 
    In (f) the bath temperature is constant, in (g) it changes as \(T(t) = T_\text{i} + 0.1\times \theta(t)\) (See text for details).
    \label{fig:4}}
\end{figure*}

To build intuition of how the potential and nonconservative forces affect the dynamics of disorder we develop a simple model based on Langevin dynamics.
We focus on the motion of the vanadium atoms relative to the high-symmetry rutile positions on the plane perpendicular to \([1, -1, 0]_\text{R}\) shown as blue and orange spheres in Fig.~\ref{fig:1}a.
For simplicity, we consider only the vanadium pairs that move primarily in this plane~\cite{Comes1974} parameterized as \(\mathbf{r}_j = (x_j, y_j)\), with \(x_j = y_j = 0\) being the average rutile positions and \(j\) a spatial index (see Methods for details).
We define a local potential centered on each high-symmetry site
\begin{equation}\label{eq:potential}
    U(x, y, t) = r(t) (x^2 + y^2) + \frac{1}{2}(x^4 + y^4),
\end{equation}
where \(r(t < 0) = -1\) and \(r(t > 0) =  0\). 
The energy units are such that \(U(\pm 1,\pm 1, t<0) = -1\) with minima at \(x = \pm 1\), \(y = \pm 1\) corresponding to the M1 positions in normalized coordinates, as shown by the contour plots in Fig. \ref{fig:4}a.
The filled circles represent the vanadium atoms in equilibrium at \(T(t<0) = T_\text{i} = 0.005\) in units relative to the depth of the potential well.
Here \(T_\text{i}\) is the initial temperature.
Photoexcitation makes \(r(t > 0) = 0\), such that the potential becomes wide with minimum at \(x = y = 0\) as suggested by observations and ab-initio calculations~\cite{Wall2018,Li2022}. 
The contours in Figs.~\ref{fig:4}b-e show \(U(x, y, t > 0)\).
The sudden change in the potential causes the distribution to broaden quickly, as shown by the purple dots in Fig~\ref{fig:4}b-e at \(t = 0.2~\mathrm{ps}\).
This broadening is associated with larger spatial fluctuations and stronger diffuse scattering~\cite{Wall2018,Budai2014}.
As we show below, this model illustrates that without additional forces the broadening is slower for lower \(T_\text{i}\).

The fast motion of the electrons in the transient metal causes additional forces not captured by the time-dependent potential in Eq.~(\ref{eq:potential}). 
We treat the electrons as an effective thermal bath characterized by a time-dependent temperature \(T(t)\) that causes stochastic and damping forces on the vanadium lattice~\cite{Head-Gordon1995,Duffy2006,Darkins2018,Darkins2018simulating,Tamm2018}.
The equations of motion for the \(\mathbf{r}_j\) in dimensionless units are
\begin{equation}\label{eq:eom}
    \frac{1}{a}\ddot{\mathbf{r}}_j =
        -\boldsymbol\nabla U(\mathbf{r}_j, t)
        - \gamma \dot{\mathbf{r}}_j
        + \boldsymbol\eta_j(t)
\end{equation}
where \(U(\mathbf{r}, t)\) is defined in Eq.~(\ref{eq:potential}),  \(\gamma = 5.3~\mathrm{ps}^{-1}\) is a phenomenological damping that must be strong enough to attenuate coherent oscillations in the Bragg peaks in Fig. \ref{fig:2}d (see also \cite{Wall2018}) and \(a = (2 \pi\, 6 \,\mathrm{THz})^2/2\) defines the units of time such that the small amplitude oscillations in the initial potential have a frequency of \(6~\mathrm{THz}\)~\cite{Wall2012}.
The term \(\boldsymbol\eta_j(t)\) represents uncorrelated random forces due to the electron bath and satisfies the fluctuation-dissipation theorem, with zero mean and variance \(\langle \eta_j^2 \rangle = 2 \gamma T(t)\), where \(T\) is given relative to the potential well depth. 
As we explain below, this stochastic term is crucial to accelerate the disorder and increase in diffuse intensity.
Importantly, the only temperature in the model, \(T(t)\), is the temperature of the electron bath. 
The vanadium lattice does not have a well-defined temperature at early times.
In all cases \(T(t < 0) = T_\text{i}\) and the lattice is initially in equilibrium.
In spite of its simplicity, as we show next, Eq.~(\ref{eq:eom}) captures the behavior of the fluctuations, diffraction and diffuse intensity reasonably well.

The timescale \(\gamma^{-1} = 0.18~\mathrm{ps}\) is associated with electronic friction where the damping constant \(\gamma\) is related to the electron-phonon coupling constant \(g\)~\cite{Head-Gordon1995,Duffy2006} (see in particular Eq. (14) in Ref. \cite{Duffy2006}). Assuming that the dominant contribution to the damping \(\gamma\) is from electronic collisions we obtain \(g = 3.7\times 10^{18}~\mathrm{W m^{-3} K^{-1}}\) which is an order of magnitude larger than that of typical metals~\cite{Qiu1992}.  
This attests to the large electron-lattice interaction of \(\mathrm{VO_2}\) and the importance of the nonconservative forces the electrons exert on the lattice.

%
%
In the following simulations we take \(T(t<0) = T_\text{i}\) with \(T_\text{i}\) the initial temperature (Fig. \ref{fig:4}a) and \(r(t>0)=0\), which changes the potential energy as depicted in Fig.~\ref{fig:4}b-e.
We consider two scenarios with different behavior in the electron bath:
1) constant temperature, \(T(t>0) = T_\text{i}\), shown in Fig.~\ref{fig:4}b-c and \ref{fig:4}f; and 
2) sudden temperature jump, \(T(t>0) = T_\text{i} + 0.1\), shown in Fig.~\ref{fig:4}d-e and \ref{fig:4}g.
We also present in Extended Data Fig. 1 an AIMD simulation which shows that the disorder slows down at low temperature in the absence of collisions, even for this complex potential derived from first principles.
Clearly, the fast disorder observed experimentally does not result from peculiarities in the potential, but from collisions with hot electrons [phenomenologically, the third term in Eq. (\ref{eq:eom})] that quickly spread the displacement distribution.

Figures~\ref{fig:4}f and \ref{fig:4}g show the simulated dynamics of the \((2.5, 2.5)\) Bragg peak (bottom panels) and diffuse intensity (top panels) integrated over the wavevector range \((0.1, 0.1) - (0.45, 0.45)\) for several initial temperatures (see Methods for details on the structure factor calculation). 
In Fig.~\ref{fig:4}f the bath temperature is constant \(T(t>0) = T_\text{i}\), and in \ref{fig:4}g it changes as \(T(t>0) = T_\text{i} + 0.1\). 
Note that \(T_\text{i} = 0.005\) corresponds to \(100~\mathrm{K}\) and \(T(t>0) = T_{\text{i}} + 0.1\) corresponds to a bath at 2000 K, in agreement with estimates of the photoexcited electronic temperature~\cite{Wall2018}.
The Bragg peaks are normalized to the initial intensity and the change in diffuse intensities $\Delta I(t)$  are shown as percentage of their respective saturation value at $t > 2$~ps.
Fig.~\ref{fig:4}f shows that while \(T_\text{i}\) does not affect the Bragg peaks significantly, in scenario 1) the diffuse intensity increases more slowly at lower \(T_\text{i}\). 
This behavior agrees qualitatively with the observed slowdown of the Debye-Waller factor (a measure of disorder) with the initial temperature\cite{Wang2020} during ultrafast melting of InSb~\cite{Lindenberg2005,Gaffney2005,Hillyard2007},
but contrasts with our observations of disorder in \(\mathrm{VO_2}\).
Note that because of the average change in symmetry between monoclinic and rutile and associated structure factor change, the Debye-Waller factor on Bragg peaks is not a good measure of disorder here.

The situation for a step in \(T(t)\) for scenario 2) in Fig.~\ref{fig:4}g is quite different.
Here, \(\Delta I(t)\) (top panel) changes with the same timescale independently of \(T_\text{i}\), indicating that the system quickly loses memory of the initial lattice fluctuations because of the additional kicks from the last term in Eq. (\ref{eq:eom}). 
Note that in this case too, the Bragg peak intensity (lower panel) is dominated by the average distortion and cannot distinguish the two scenarios (compare lower panels in Fig.~\ref{fig:4}f and \ref{fig:4}g). 

These results indicate that electronic degrees of freedom can exert additional forces on the lattice that are not captured by the potential energy surfaces often used in phenomenological\cite{Sun2020}, or first-principles-based~\cite{Wall2018,Li2022,Liu2022} descriptions of ultrafast structural dynamics.
We showed that these forces have a strong influence in the lattice dynamics of \(\mathrm{VO_2}\) and they most strongly affect the fluctuations rather than the average motion and thus their characterization requires measurements beyond those of diffraction~\cite{Wall2018,Trigo2013,Stern2018,Chase2016}. 
Materials control by light pulses will require not only more sophisticated measurements of structural dynamics, but also a fundamental description of the coupled electron and lattice degrees of freedom beyond the Born-Oppenheimer approximation~\cite{Picano2022}.

\section{\label{sec:Acknowledgements} Acknowledgements}
G.A.P.M., Y.H., V.K., D.A.R., S.T. and M.T. were supported by the US Department of Energy, Office of Science, Office of Basic Energy Sciences through the Division of Materials Sciences and Engineering under Contract No. DE-AC02-76SF00515. 
A. A. C. work was supported by the Center for Non-Perturbative Studies of Functional Materials Under Non-Equilibrium Conditions (NPNEQ) 
funded by the Computational Materials Sciences Program of the U.S. Department of Energy, Office of Science, Basic Energy Sciences, Materials Sciences and Engineering Division
and performed under the auspices of the U.S. Department of Energy by Lawrence Livermore National Laboratory under Contract DE-AC52-07NA27344.
OD was supported by the US Department of Energy (DOE), Office of Science, Basic Energy Sciences, Materials Sciences and Engineering Division, under Award No. DE-SC0019978.
ASJ, EP, LV and SW were funded through the European Research Council (ERC) under the European Union’s Horizon 2020 Research and Innovation Programme (Grant Agreement No. 758461) and PGC2018-097027-B-I00 project funded by MCIN/ AEI /10.13039/501100011033/ FEDER “A way to make Europe” and CEX2019-000910-S [MCIN/ AEI/10.13039/501100011033], Fundació Cellex, Fundació Mir-Puig, and Generalitat de Catalunya (AGAUR Grant No. 2017 SGR 1341, CERCA program). 
ASJ and EP acknowledge support from the Marie Skłodowska-Curie grant agreement no. 754510 (PROBIST). 
ASJ acknowledges support of a fellowship from ”la Caixa” Foundation (ID 100010434), fellowship code LCF/BQ/PR21/11840013 and the Agencia Estatal de Investigacion (the R\&D project CEX2019-000910-S, funded by MCIN/ AEI/10.13039/501100011033, Plan National FIDEUA PID2019-106901GB-I00, FPI).
TK acknowledges support from JSPS KAKENHI (Grant Numbers JP19H05782, JP21H04974, and JP21K18944).
Ultrafast x-ray measurements were performed at BL3 of SACLA with the approval of the Japan Synchrotron Radiation Research Institute (JASRI) (Proposal Nos. 2019A8038 and 2019B8075).
Preliminary x-ray characterization was performed at the Stanford Synchrotron Radiation Lightsource (SSRL). Use of the SSRL is supported by the US Department of Energy, Office of Science, Office of Basic Energy Sciences under Contract No. DE-AC02-76SF00515. 

\section{\label{sec:AuthorContributions} Authors contributions}
M.T., S.W. and O.D. conceived the initial experiment. M.T. supervised the project. G.A.P.M., Y.H., A.J., T.K, V.K., E.P., D.A.R., S.T., L.V., S.W. and M.T. performed the experiment at SACLA. G.A.P.M. analyzed the data. M.T. and A.A.C. developed the model, and M.T., G.A.P.M. and S.Y. performed the simulations. M.T. and A.A.C. wrote the initial draft with feedback from all the authors.

\section{Competing Interests Statement}
The authors declare no competing interests.



\section{\label{sec:Methods} Methods}

\subsection{Experimental details}

The ultrafast x-ray scattering measurements were performed at the Spring-8 Angstrom Compact free electron LAser (SACLA) facility in Japan using x-ray pulses with \(10~\mathrm{fs}\) duration and \(12~\mathrm{keV}\) photon energy (probe).
The sample consisted of a single-crystal of \(\mathrm{VO_2}\) with the sample normal along the rutile \([1,1,0]_R\) direction \cite{Wall2018}.
The sample was photo-excited with laser pulses (pump) centered at a wavelength of \(800~\mathrm{nm}\), with duration of \(40~\mathrm{fs}\).  
The arrival time of pump and probe were measured and corrected on a shot-by-shot basis giving an overall time resolution of \(50~\mathrm{fs}\) full width at half maximum (FWHM).
We used a grazing geometry with the incident x-ray beam making an angle of \(0.5~\mathrm{deg}\) with respect to the sample surface to match the penetration depth of the optical pulse.
The pump was nearly co-linear with the probe (\(5~\mathrm{deg}\)).
Scattered x-ray photons were captured by the MultiPort Charged Coupled Device (MPCCD) detector, which we placed \(96~\mathrm{mm}\) away from the sample to cover a large section of reciprocal space.
A nitrogen cryo-stream was used to cool the sample to \(100~\mathrm{K}\).

\subsection{Model of disorder dynamics}

The two-dimensional model is based on the dynamics of particles coupled to a Langevin bath.
The vanadium atoms evolve according to Eq.~(\ref{eq:eom}) in the main text. 
The coordinates are normalized such that the positions at the minima of the potential are \(\mathbf{r} =  (\pm 1,\pm 1)\). 
Starting with a staggered initial condition \(\mathbf{r}_0 = (1, 1) (-1)^{i+j}\) with \(i\) and \(j\) the indices of the vanadium in the plane, we run the simulation using an Euler-Maruyama scheme for \(4~\mathrm{ps}\) to obtain an initial thermal ensemble.
In Fig.~\ref{fig:4}a we show a snapshot of this distribution thermalized at a temperature \(T = 0.005\) in units of the potential well depth.
The time dependent excitation is modeled as a change in the coefficient from \(r(t < 0) = -1\) to \(r(t \geq 0) = \exp(-i/x_0)\theta(t)-1\), which accounts for a spatially decaying excitation along the sample depth (index \(i\)), necessary to reproduce the Bragg peak contrast observed experimentally due to a mismatch of the x-ray penetration depth.
We find that \(x_0 = 1/3\) of the x-ray penetration depth reproduced the observed \(0.5\) contrast in the Bragg peaks, however this does not affect the resulting time-scales.
For simplicity we assume a square lattice with lattice constant \(a_\text{R} = 5.78~\mathrm{\text{\AA}}\). 

We compute the structure factor as \(S(\mathbf{Q},t) \propto \langle|\sum_j e^{i \mathbf{Q}\cdot (\mathbf{R}_j +  \mathbf{r}_j(t))}|^2 \rangle\), where \(\mathbf{R}_j\) are the equilibrium positions in the high-symmetry structure, $\mathbf{Q}$ is the momentum transfer in reciprocal lattice units (rlu), and the average is over realizations of the random forces.
Wavevectors in rlu are given in the reciprocal lattice of the high-symmetry structure such that half-integers represent Bragg peaks of the distorted 2D structure analogous to the M1 structure.

\section{\label{sec:DataAvailability} Data Availability Statement}
Raw data were generated at the SACLA large-scale facility. Derived data supporting the findings of this study are available from the corresponding author upon request.

\section{\label{sec:CodeAvailability} Code Availability Statement}
Computer codes used in this study are available from the corresponding author upon request.

\end{document}